# Investigation of Flow Characteristics inside a Dual Bell Nozzle with and without Film Cooling

Mayank Verma, Nitish Arya, Ashoke De[*]

*Indian Institute of Technology Kanpur, Kanpur, Uttar Pradesh, 208016, India*

**Abstract**

In this study, we perform a two-dimensional axisymmetric simulation to assess the flow characteristics and understand the film cooling process in a dual bell nozzle. The secondary stream with low temperature is injected at three different axial locations on the nozzle wall, and the simulations are carried out to emphasize the impact of injection location (secondary flow) on film cooling of the dual bell nozzle. The cooling effect is demonstrated through the temperature and pressure distributions on the nozzle wall or, in-turn, the separation point movement. Downstream of the injection point, the Mach number and temperature profiles document the mixing of the main flow and secondary flow. The inflection region is observed to be the most promising location for the injection of the secondary flow. We have further investigated the effect of Mach number of the secondary stream. The current study demonstrates that one can control the separation point in a dual bell nozzle with the help of secondary injection (Mach number) so that an optimum amount of thrust can be achieved.

## I. Introduction

The next generation space transportation systems demand an increase in operational efficiency and a reduction in the Earth to Orbit launch costs. Since there is intense competition for space transportation, the rocket engines should be designed to deliver high performance at a lower system complexity. Most of the current launcher systems, like the European launcher Ariane 5 utilize parallel staging [1]. The main engine has to operate over a wide range of altitudes resulting in the off-design operation. The off-design operation of the engine produces a significant performance loss in the conventional bell-type nozzles, where the flow over or under expands due to the fixed exit area ratio. This may sometimes result in 10% or more loss in the specific impulse as compared to an ideal contour nozzle [2]. To reduce this performance loss, various concepts of altitude adaptive nozzles have been proposed in the literature, among which the most promising concept is the dual bell nozzle [3-10].

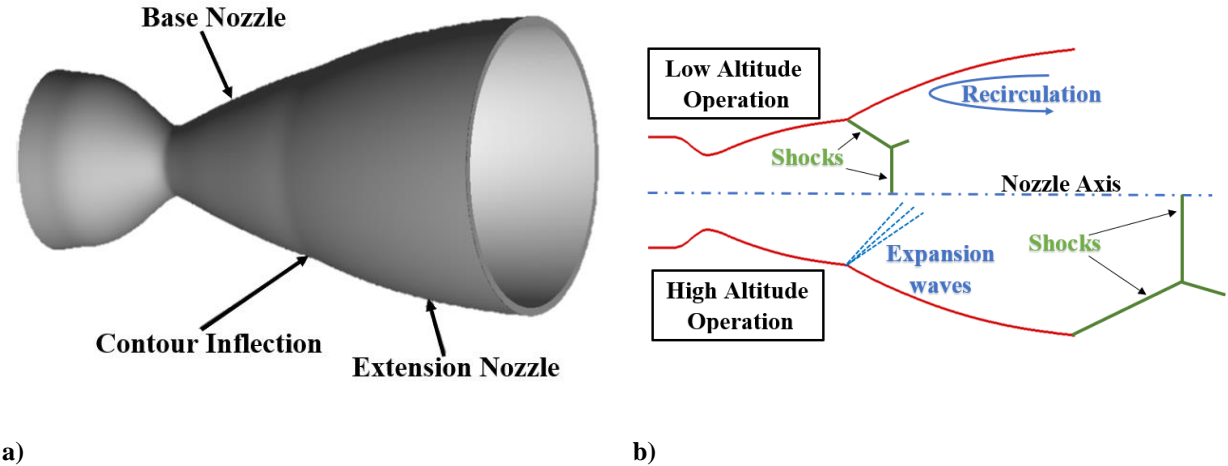

**Fig. 1. a)** Schematic of a dual bell nozzle **b)** Operating modes of a dual bell nozzle

A dual bell nozzle combines two differently designed conventional nozzles into one nozzle. The first nozzle with a relatively low expansion ratio is called "Base nozzle", while the second nozzle with the larger expansion ratio is commonly known as "Extension nozzle" in the literature of nozzles [10-11]. Figure 1 a) depicts schematic contours of the dual bell nozzle. Essentially, the dual bell nozzle is an altitude-adapting nozzle which does not contain any movable parts but allows one-step change from low-altitude operation mode to the high-altitude operation mode. Due to this attractive property, it has recently gained scientific attention. For the low altitude operation mode, the main flow from the nozzle inlet fills the base nozzle only while the flow separation point stabilizes at the inflection point (Fig. 1 b). Contrarily, for the high-altitude operation mode, the main flow from the nozzle inlet fills the base nozzle and the extension nozzle as well, while the separation point shifts at the lip of the extension nozzle. During the ascent of the rocket from sea-level to space, the working mode transition from the low altitude operation mode to high altitude operation mode takes place within the dual bell nozzle [11-17]. In particular, the fluid dynamics parameter purely governs this transition from the full-flowing base nozzle to the full flowing extension nozzle with the altitude change, i.e. nozzle pressure ratio, NPR, which is the ratio of the combustion chamber pressure to the ambient pressure. However, the nozzle thrust decreases due to the high side loads during such transition [11, 13, 16, 18-19].

In the case of clustered engines, the NPR for each engine at the same altitude may be different, which, in turn, results in different transition altitudes for each engine. Because of the variation in the transition altitude of the different engines, there exists a thrust imbalance in each engine, thereby severely influencing the vehicle trajectory. Thus, active control of the transition is of vital importance for dual bell nozzles. Proschanka et al. [12], Martelli et al. [20], and Arnold et al. [21-22] investigated the effectiveness of the film cooling and concluded that the injection of the secondary flow into the nozzle provides an active control on the transition without any mechanical activation. In most of the studies of the film cooling effectiveness in the dual bell nozzle [12, 20-23], the injection location for the secondary stream is randomly selected without giving any justification. No comparison study is performed to investigate the effect of the injection location until now. Furthermore, most of the studies have considered the main flow to be at atmospheric temperature and the secondary flow to be at a relatively lower temperature than the atmospheric temperature. Thus, we have made an attempt to assess the film cooling phenomenon inside the dual bell by keeping the main flow at a temperature of 2842K (adopted from a real engine test data) and the secondary flow at 500 K. The properties related to the main flow at the temperature of 2842 K are mentioned in the subsequent sections of the paper.

In the current work, we have invoked ANSYS FLUENT v18.0 as CFD (Computational Fluid Dynamics) solver to investigate the effect of the injection location of the secondary stream in a dual bell nozzle by injecting at three different axial locations on the nozzle wall viz. at nozzle throat, at the base nozzle and at inflection region. The inflection region is found to be the most favorable location for secondary stream injection. Both the pressure and temperature distributions over the nozzle walls confirm the above observation. Furthermore, the analysis of the boundary layer profile and near-wall temperature profile demonstrates the mixing of hot and cold flow in the dual bell nozzle. Finally, we have also demonstrated the effect of secondary stream Mach number on the separation point, which exhibits that it is possible to control the working operation mode of the dual bell nozzle with the help of secondary flow Mach number. Thus, the present study provides a comprehensive analysis of the film cooling inside a dual bell nozzle and forms a basis for the future researchers to properly select the parameters affecting the film cooling behavior in dual bell nozzle.

## II. Numerical Setup

### II.A. Geometry and Meshing Details

*Nozzle Geometry:* The dual bell nozzle considered in this study is extracted from one of the pairs of base nozzles and extension nozzles, extensively studied by Tomita et al. [6]. Following the naming rules of Tomita et al. [6], the B-TO-SLS/E-TO-LLS is considered for this study. "B-" at the head means base nozzle as well as "E-" means extension nozzle. "TO-" denotes the design method of the nozzle, i.e. parabolic. "SLS" and "LLS" at the last group represents the configuration of the nozzle, namely, the nozzle length, initial expansion angle, and the nozzle exit angle in the sequential order. Each character, "S", "M", and "L' denotes small, medium, and large, respectively. The base nozzle is a parabolic thrust optimized contoured nozzle with small nozzle length, with a large initial expansion angle and with a small nozzle exit angle (B-TO-SLS). Whereas, the extension nozzle is designed to be a parabolic nozzle with long nozzle length, with a large initial expansion angle and with a small nozzle exit angle (E-TO-LLS). Figure 2 illustrates the contours of the base and extension nozzle, and Table 1 provides the design parameters.

**Table 1**
Design parameters of dual bell nozzle

| Nozzle | Design Method | Nozzle Length/Throat Radius | Initial Expansion Angle, degree | Nozzle Exit Angle, degree |
|---|---|---|---|---|
| Base Nozzle | Parabolic | 4 | 25 | 5 |
| Extension Nozzle | Parabolic | 6.25 | 30 | 5 |

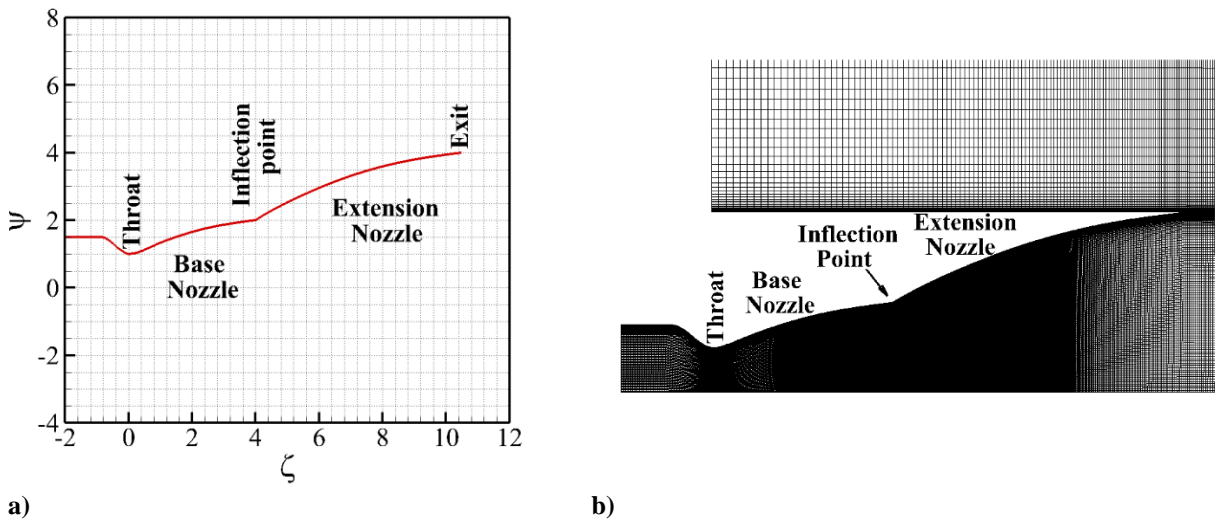

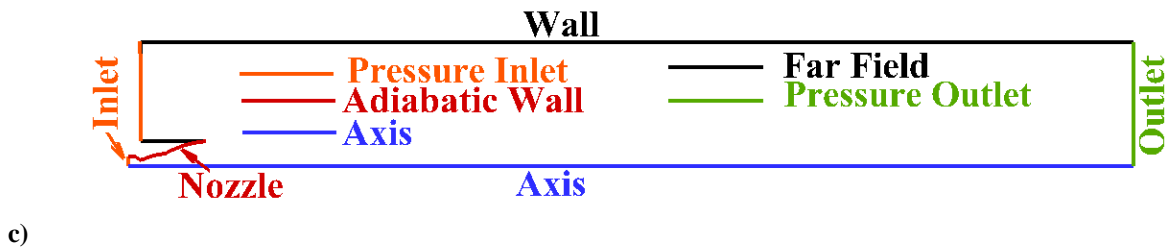

c)

**Fig. 2 a)** Nozzle test article contour [2] **b)** Meshing view of the nozzle **c)** Computational domain with corresponding boundary conditions used for the calculations.

*Computational Domain:* To develop the appropriate NPR across the nozzle inlet and the nozzle outlet, the computational domain is extended 100 times the throat radius ($r_{th}$) in the axial direction and 40 times $r_{th}$ in the radial direction after the nozzle outlet. The computational domain is further extended in the upstream direction until the nozzle throat from the nozzle outlet to capture the atmospheric inflow. A multi-block structured mesh is generated using ICEM-CFD v18.0 software. The nozzle wall is appropriately resolved by keeping the $y^+$ value lower than one throughout the nozzle wall to capture the laminar sub-layer region accurately. The nozzle contour, the corresponding mesh, and the computational domain are sketched in Fig. 2, where ψ represents the non-dimensional ordinate (y/$r_{th}$), and ζ is the non-dimensional abscissa (x/$r_{th}$).

*Film Cooling Geometry:* To study the film cooling effects of the secondary flow injection, it is injected parallel to the nozzle axis at three different locations on the nozzle wall viz. at nozzle throat ($\zeta = 0$), at the base nozzle ($\zeta = 2$) and at the inflection point ($\zeta = 4$). The injection is made by shifting the post-secondary-flow-inlet geometry in the normal direction by a slot height, $s = 1\ mm$. Figure 3 exhibits the modified nozzle geometry for film cooling injection and corresponding meshing views for different injection locations.

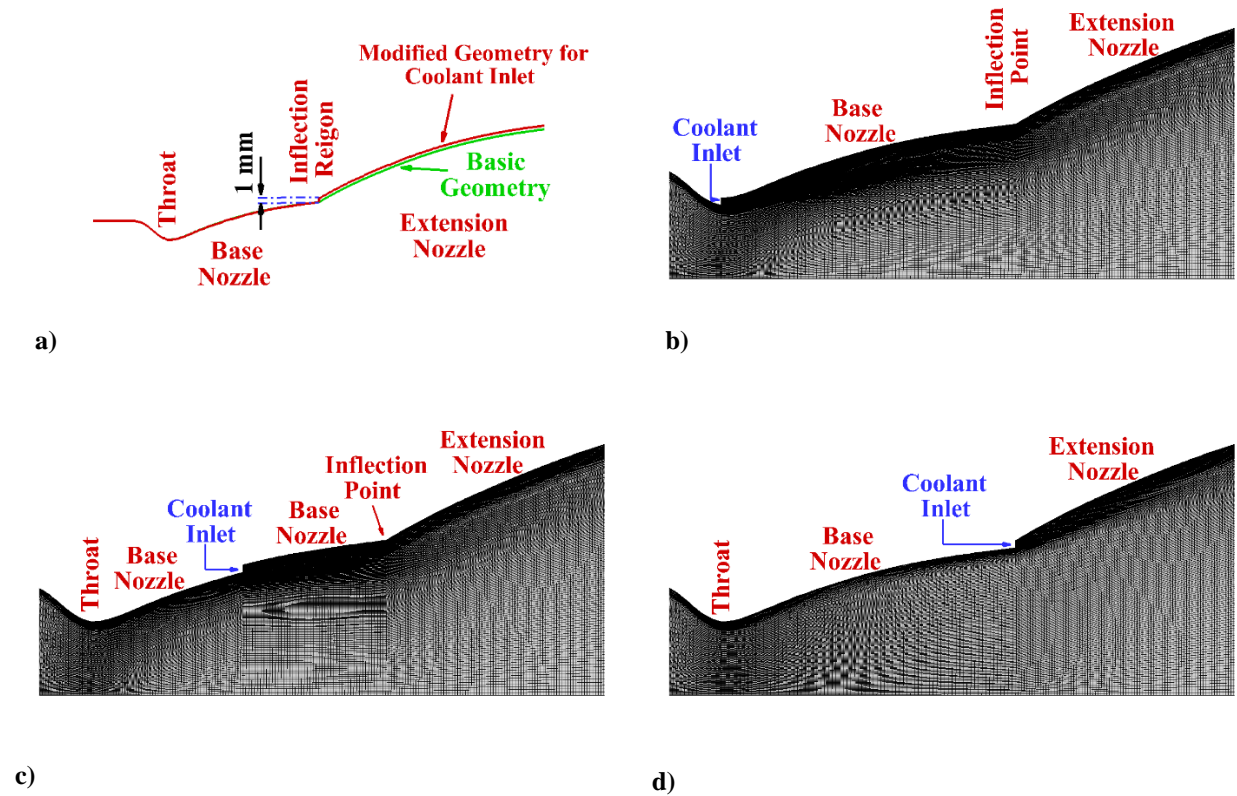

**Fig. 3. a)** Modified nozzle geometry for secondary flow inlet; Meshing view for secondary flow inlet at **b)** nozzle throat **c)** base nozzle **d)** inflection point

## II.B. Governing Equations and Solver Details

We have performed a two-dimensional axisymmetric study by solving the Favre averaged, steady equations for mass, momentum, and energy conservation along with turbulence quantities [24-25]. The following equations are solved using the Finite Volume method along with the perfect gas equation.

Continuity:
$$\frac{\partial}{\partial x_i}(\bar{\rho}\tilde{u}_i) = 0 \qquad (1)$$

Momentum:
$$\frac{\partial}{\partial x_j}(\bar{\rho}\tilde{u}_i\tilde{u}_j) + \frac{\partial}{\partial x_i}(\bar{p}) - \frac{\partial}{\partial x_j}\left((\mu + \mu_t)\frac{\partial \tilde{u}_i}{\partial x_j}\right) = 0 \qquad (2)$$

Energy:
$$\frac{\partial}{\partial x_j}(\bar{\rho}\tilde{u}_j\tilde{E}) + \frac{\partial}{\partial x_j}\left(\tilde{u}_j\left(-\bar{p}I + \mu\frac{\partial \tilde{u}_i}{\partial x_j}\right)\right) - \frac{\partial}{\partial x_j}\left(\left(\Gamma + \frac{\mu_t C_p}{Pr_t}\right)\frac{\partial \tilde{T}}{\partial x_j}\right) = 0 \qquad (3)$$

Equation of State:
$$\bar{p} = \bar{\rho}R\tilde{T} \qquad (4)$$

The symbols (^) and (~) represent the time averaging through Reynolds decomposition and density-weighted time averaging (Favre averaging), respectively. In the above equations, $\rho$ is the density, $u_i$ is the velocity vector, $T$ is the temperature, $p$ is the pressure, $E = e + u_i^2/2$ is the total energy, where $e = h - p/\rho$ is the internal energy, and $h$ is enthalpy. The fluid properties $\mu$ and $\Gamma$ are respectively the viscosity, and the thermal conductivity, while $\mu_t$, and $Pr_t$ are the turbulent eddy viscosity, and turbulent Prandtl number, respectively. The turbulent Prandtl number is taken as 0.85, and the dynamic viscosity is evaluated through Sutherland's law, as described below. The $\mu_t$ is evaluated by complementing the above set of equations along with the transport equations of turbulence quantities. In the present study, the SST k-ω turbulence model is invoked to close the above set of equations. The eddy viscosity $\mu_t$ is modeled as:

Eddy Viscosity:
$$\mu_t = \frac{\rho a_1 k}{max(a_1\omega, \Omega F_2)} \quad with\ a_1 = 0.35 \qquad (5)$$

Where $\Omega$ is the vorticity magnitude, $F_1$ and $F_2$ is a blending function given as:

$$F_2 = \tanh\left\{\left(max\left[2\frac{\sqrt{k}}{0.09\omega y}, \frac{500\mu}{\rho y^2 \omega}\right]\right)^2\right\} \qquad (6)$$

$$F_1 = \tanh\left\{\left(min\left[max\left\{\frac{\sqrt{k}}{0.09\omega y}, \frac{500\mu}{\rho y^2 \omega}\right\}, \frac{4\rho\sigma_{\omega_2}k}{CD_{k\omega}y^2}\right]\right)^4\right\} \qquad (7)$$

Where,
$$CD_{k\omega} = \min\left(2\rho\sigma_{\omega_2}\frac{1}{\omega}\frac{\partial k}{\partial x_j}\frac{\partial \omega}{\partial x_j}, 10^{-20}\right) \qquad (8)$$

| Turbulent kinetic energy: | $$\frac{\partial}{\partial x_j}\left(\rho k u_j - (\mu + \mu_t \sigma_k)\frac{\partial k}{\partial x_j}\right) = \tau_{ij}\frac{\partial u_i}{\partial x_j} - \beta^* \rho k \omega$$ | (9) |

| Specific Dissipation rate: | $$\frac{\partial}{\partial x_j}\left(\rho \omega u_j - (\mu + \mu_t \sigma_\omega)\frac{\partial \omega}{\partial x_j}\right) = P_\omega - \beta \rho \omega^2 + 2(1 - F_1)\rho \sigma_{\omega_2}\frac{1}{\omega}\frac{\partial k}{\partial x_j}\frac{\partial \omega}{\partial x_j}$$ | (10) |

$$P_\omega = \rho \frac{\gamma}{\mu_t} \tau_{ij} \frac{\partial u_i}{\partial x_j} \tag{11}$$

$$\tau_{ij} = \mu_t \left(2 S_{ij} - \frac{2}{3}\frac{\partial u_k}{\partial x_k}\delta_{ij}\right) - \frac{2}{3}\rho k \delta_{ij} \tag{12}$$

The coefficients are blended through the blending function: $\varphi = \varphi_1 F_1 + (1 - F_1)\varphi_2$, and given as $\sigma_{k1} = 0.85$, $\sigma_{k2} = 1.0, \sigma_{\omega 1} = 0.5, \sigma_{\omega 2} = 0.856, \gamma_1 = 0.5532, \gamma_2 = 0.44$. Further details of the above equations can be found in the Refs. [26, 27, 28].

The solver, i.e. ANSYS FLUENT v18.0, uses a finite volume approach to discretize the compressible Navier-Stokes equations and employs a density-based formulation solves the governing equations of continuity, momentum, and energy in a coupled fashion. Spatial discretization of the convective fluxes is achieved by the second-order upwind scheme and evaluated using a flux-vector splitting scheme following the Advection Upstream Splitting Method, termed AUSM. A second-order central scheme is used to discretize the viscous fluxes. Simulations are performed by setting the Courant-Friedrich-Lewy number equal to 0.5.

The dependence of molecular viscosity on temperature is accounted for by incorporating Sutherland's law [29] and recast as:

$$\mu = \mu_o \left(\frac{T}{T_o}\right)^{\frac{3}{2}} \frac{T_o + S}{T + S} \tag{13}$$

Where $\mu$ is the viscosity in kg/m-s, $T$ is the static temperature in K, $\mu_o$ is the reference value of viscosity in kg/m-s, $T_o$ is the reference temperature in K, and $S$ is the effective temperature in K. For air at moderate temperatures and pressures, $\mu_o = 1.716 \times 10^{-5}$ kg/m-s, $T_o = 273.11\ K$, and $S = 110.56\ K$. The molecular weight of the main flow is 21.252 kg/kmol, and the thermal conductivity is 0.35797 W/m-K. Piecewise polynomial of 8$^{th}$ order is used to account for the dependence of specific heat capacity at constant pressure (C$_p$) on temperature.

$$for\ T_{min,1} \leq T \leq T_{max,1}:\ C_p(T) = A_1 + A_2 T + A_3 T^2 + \cdots \tag{14}$$

$$for\ T_{min,2} \leq T \leq T_{max,2}:\ C_p(T) = B_1 + B_2 T + B_3 T^2 + \cdots \tag{15}$$

Where $T_{min,1} = 100\ K$, $T_{max,1} = 1000\ K$, $T_{min,2} = 1000\ K$, $T_{max,2} = 3000\ K$, $A_1$, $A_2$, $A_3$... and $B_1$, $B_2$, $B_3$... are the constants. The values of these constants are given below in Table 2.

**Table 2**
Value of constants used in polynomial for specific heat capacity at constant pressure

| Suffix | 1 | 2 | 3 | 4 | 5 | 6 | 7 | 8 |
|---|---|---|---|---|---|---|---|---|
| A | 1161.482 | -2.36881 | 0.0148 | -5.03e-05 | 9.92e-08 | -1.1e-10 | 6.54e-14 | -1.5e-17 |
| B | -7069.814 | 33.70605 | -0.0581 | 5.42e-05 | -2.93e-08 | 9.23e-12 | -1.5e-15 | 1.11e-19 |

**Table 3**
Boundary conditions used for the computations

| Boundary | Type | Condition |
|---|---|---|
| Nozzle Inlet | Pressure Inlet | $p_0$ = NPR × $p_a$, $T_0$ = 300 K (for cold flow), 2842 K (for hot flow) |
| Nozzle Wall | Wall | No-slip and adiabatic |
| Axis | Axis | - |
| Domain Inlet | Pressure Inlet | $p_0$ = 101325 Pa and $T_0$ = 300 K |
| Secondary Flow Inlet | Velocity Inlet | $V_c$ = Mach No. * speed of sound, $P_c$ = 101325 Pa, $T_c$ = 500 K |
| Domain Walls | Wall | Slip and adiabatic |
| Domain Outlet | Pressure Outlet | $p_a$ = 101325 Pa |

**II.C. Boundary Conditions**

A fixed value of pressure is applied at the nozzle inlet to develop the flow inside the dual bell nozzle for the desired NPR. No-slip boundary condition is applied to the nozzle wall. Nozzle wall and the domain walls are treated as adiabatic. We have applied the pressure inlet boundary condition, with $p_a = 101325\ Pa$, at the left side of the domain, to allow the atmospheric inflow. At the domain outlet, we have used the pressure-outlet boundary condition with the value of atmospheric pressure. The secondary flow is injected using the velocity inlet boundary condition at the secondary-flow-inlet with the fixed temperature value. Characterization of turbulence quantities at the gas inflows (ambient and nozzle gas supply) is unknown for the tests. Hence, we have considered a low turbulence level with a turbulence intensity of 2% in these computations at the nozzle inlet. Figure 2 c) and Table 3 reports the boundary conditions used for the computations.

| Grid | Number of nodes | Inside nozzle |
|---|---|---|
| Grid-3 | 108,854 | 200×105 |
| Grid-2 | 247,500 | 300×160 |

| Grid-1 | 387,000 | 375×200 |
|---|---|---|

## III. Results

### III.A. Grid Independence Study, Numerical Validation, and Verification

The flow inside a dual bell nozzle is supersonic and involves shock interactions. Thus, it is essential for the numerical model to correctly capture the flow features of the supersonic flow with the supersonic secondary injection. The present numerical model employs the k-ω SST turbulence model with the modified coefficients for separated flows, proposed by Allamaprabhu et al. [28]. A mesh is generated both inside the nozzle and the domain using ICEM-CFD v18.0. Grids are refined up to the $y^+$ value of 1 at the inner wall of the nozzle where the high gradients are believed to occur. We have conducted a grid independence study on three sets of grids having 3.87 million, 2.47 million, and 1.08 million cells to validate the accuracy of the numerical model. A sufficient number of nodes are present inside the nozzle for all three sets of grids, as given in Table 4. Figure 4 a) depicts the Mach number distribution over the nozzle axis for all three sets of meshes.

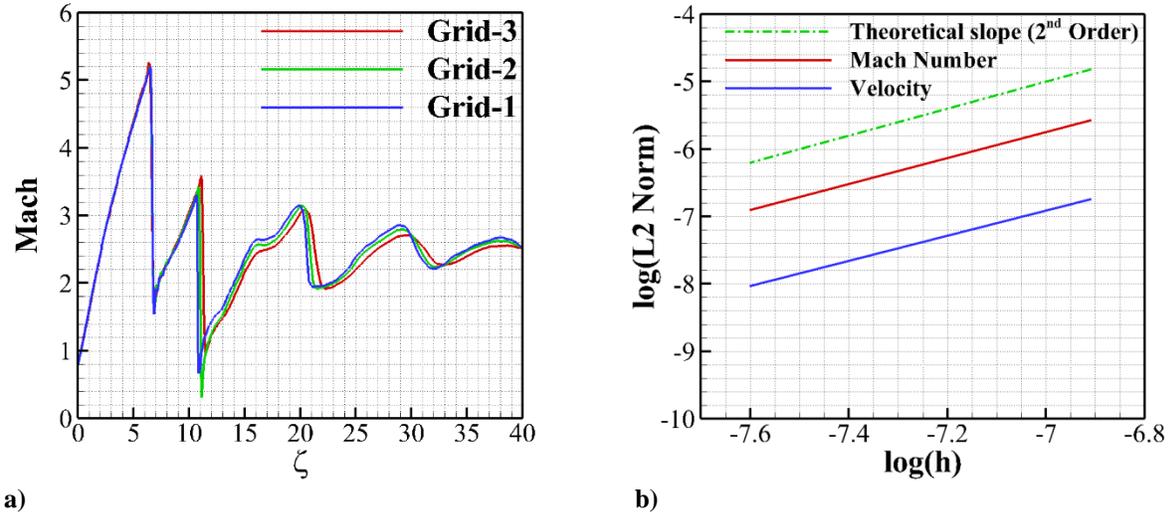

**Fig. 4. a)** Grid Independence study at NPR 25 **b)** log-log plot of L2 norm against the grid spacing h

For further analysis, we take Grid-2 as the base grid and approximate the error in Grid-1 (fine-grid) compared to Grid-2 from Richardson error estimator, defined by,

$$E_1^{Fine} = \frac{\epsilon}{1 - r^o} \qquad (16)$$

Error in Grid-3 (coarse grid), compared to the solution of Grid-2, is approximated by coarse-grid Richardson error estimator, which is defined as,

$$E_2^{Coarse} = \frac{r^o \epsilon}{1 - r^o} \quad (17)$$

Where the error is calculated from the solution of two consecutive grids by,

$$\epsilon = \frac{(f_2 - f_1)}{f_1} \quad (18)$$

Grid-Convergence Index (GCI), is calculated, which provides a uniform measure and also accounts for the uncertainty in the Richardson error estimator. GCI for fine-grid and coarse-grid is given as [30-31],

$$GCI_{Fine} = F_S \left| E_1^{Fine} \right| \quad (19)$$

$$GCI_{Coarse} = F_S \left| E_2^{Coarse} \right| \quad (20)$$

GCI is calculated for the grid refinement and coarsening of the base grid (Grid-2). A second-order accuracy ($o = 2$) with the factor of safety of 1.25 is considered for the best estimation of the grid convergence relating to a 50% grid refinement (coarsening) [32]. Mach number at the nozzle axis is evaluated for grid convergence study and tabulated in Table 5. Error estimation by Richardson extrapolation and GCI for the fine grid is relatively low compared to the coarse grid.

**Table 5**
Richardson error estimation and Grid-Convergence Index for three sets of grids

| | $r_{32}$ | $r_{21}$ | o | $\epsilon_{32}$ ($10^{-2}$) | $\epsilon_{21}$ ($10^{-2}$) | $E_{32}^{Coarse}$ | $E_{21}^{Fine}$ | $GCI_{Coarse}$ (%) | $GCI_{Fine}$ (%) |
|---|---|---|---|---|---|---|---|---|---|
| Mach number at $\zeta = 20.08$ | 1.5 | 1.25 | 2 | 2.665 | 0.155 | -0.04797 | -0.00275 | 5.996 | 0.344 |

Moreover, the L$_2$ norm is also calculated to verify the accuracy of the chosen numerical schemes. Here, we have considered Grid-1 (fine mesh) as a reference for evaluating the L$_2$ norm. Figure 4 b) documents the L$_2$ norm of the discretization error as a function of grid spacing, h, and defined as:

$$L_2 = \sqrt[2]{\left( \sum_{i=1}^{N} |\epsilon|^2 \right) / N} \quad (21)$$

Where N is the total number of nodes considered for calculation of the norm. From Fig. 4 b), the chosen schemes are found to be good enough as it nicely corroborates with the theoretical order of accuracy (theoretical slope of order is 2 and the slope obtained in the present simulation is ~ 1.92). Hence, Grid-2 has been chosen for detailed calculations reported in the following sub-sections.

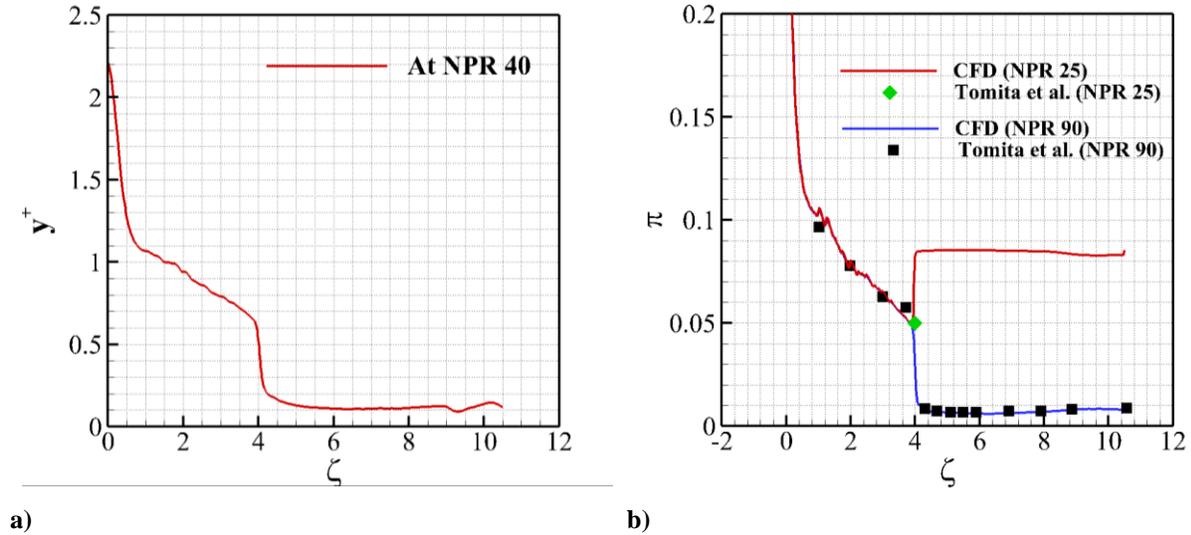

**a)** **b)**

**Fig. 5. a)** Distribution of $y^+$ value over nozzle wall at NPR 40 for Grid-2 **b)** Validation with the experimental result

In addition to the verification of numerical schemes and grid, we have validated the numerical results for both the operating modes (where NPR 90 corresponds to high altitude mode and NPR 25 corresponds to low altitude mode) using Grid-2 against the experimental results of Tomita et al. [6]. Tomita et al. [6] performed the cold flow tests on various NPRs and predicted the sideload characteristics and transition characteristics of the dual bell nozzle [10-11]. The predicted numerical results are found to be in good agreement with the experimental results for both the operating modes as depicted in Fig. 5 b), where Л is the nozzle wall pressure (p) normalized with the combustion chamber pressure ($p_c$) at the nozzle inlet, ($p/p_c$).

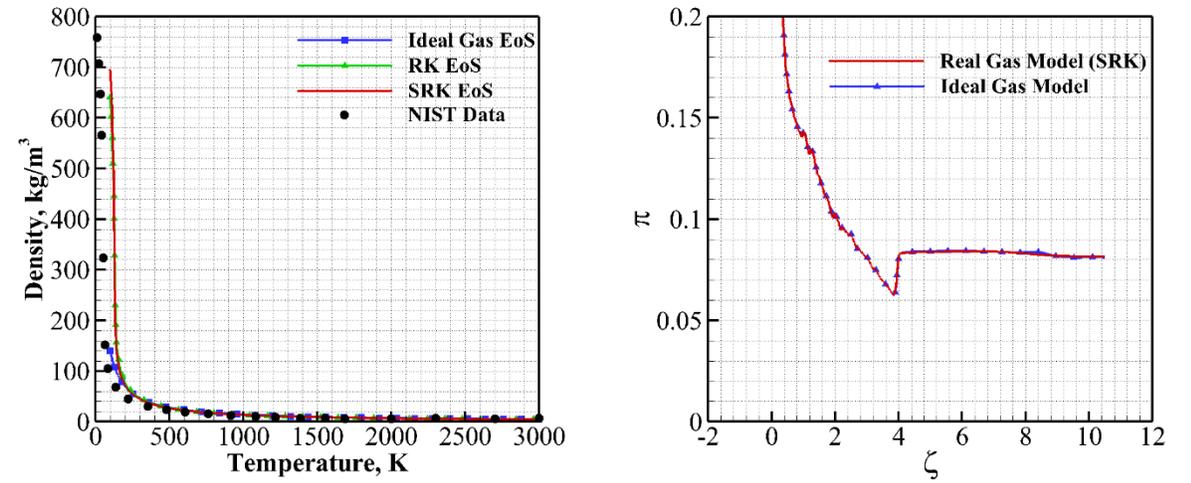

**a)** **b)**

**Fig. 6.** Comparison of the Ideal gas equation of state and different real gas equation of state. **a)** density variation with temperature **b)** normalized wall pressure profiles at NPR 25

To support the use of ideal gas equation for the calculation of the density of the high-temperature gas at the inlet of the nozzle, a comparison is made for the density variation with the temperature in Figure 6 a) at NPR 40 (as this is the NPR used for assessment of the film cooling). Noticeably, both the ideal gas equation of state and different real gas equation of state predict the same value of the density under our working range of the temperatures ($500K < T < 3000K$). Further, wall pressure profiles are compared between Soave-Redlich-Kwong, i.e. SRK EoS, and ideal gas EoS [33-34] and validate with NIST data [35]. Figure 6 b) reports the wall pressure behavior at NPR 25 for SRK EoS and ideal gas EoS. As found, both models predict the same pressure distribution over the nozzle wall. Hence, the selection of ideal gas EoS is sufficient for the calculation of the density.

*<u>Numerical Validation of Supersonic Film Cooling Model</u>*

Since the secondary flow is injected to cool the nozzle and both the main flow and the secondary flow are supersonic; hence, it is quite essential for the numerical model to correctly capture the flow features of the supersonic flow with the secondary injection. Thus, the validation study is performed with the experimental and numerical studies by Aupoix et al. [36]. Aupoix et al. [36] had extensively studied supersonic injection using Nitrogen and air mixture with $T_{secondary} = 125\ K$ at $M_{secondary} = 2$ as a coolant at the bottom of the main flow with dry air with $T_{main} = 320\ K$ at $M_{main} = 2.78$. Figure 7 a) illustrates the computational domain used for this validation study. The total pressure of the flow was 800 mbar. The computational domain is a 2D planar geometry with a height of 154 mm, length of 1500 mm, lip thickness of 2 mm, and the secondary inlet height as 5.6 mm. A structured mesh with $2 \times 10^5$ cells are used (after grid independence check) for the numerical study.

In the available literature of film cooling, people have used different algebraic turbulence models like Baldwin-Lomax [37], various forms of k-ε models [38], and k-ω models [39] for nozzle flow. But these models over-predict the mixing and turbulent length scales for the flow with the secondary injection. k-ω SST model blends the k-ω model (at the near-wall regions) with the k-ε model (at the regions away from the wall) using a blending function. Our film-cooling model validation study also depicts the same. Figure 7 b) reports the adiabatic wall temperature distribution along the length of the plate. It is further compared with the experimental data as well as with other turbulence models studied by Aupoix et al. [36]. The numerical results with the adopted computational methodology are in good agreement with the experimental results and also predict the better mixing and cooling as compared to other numerical models.

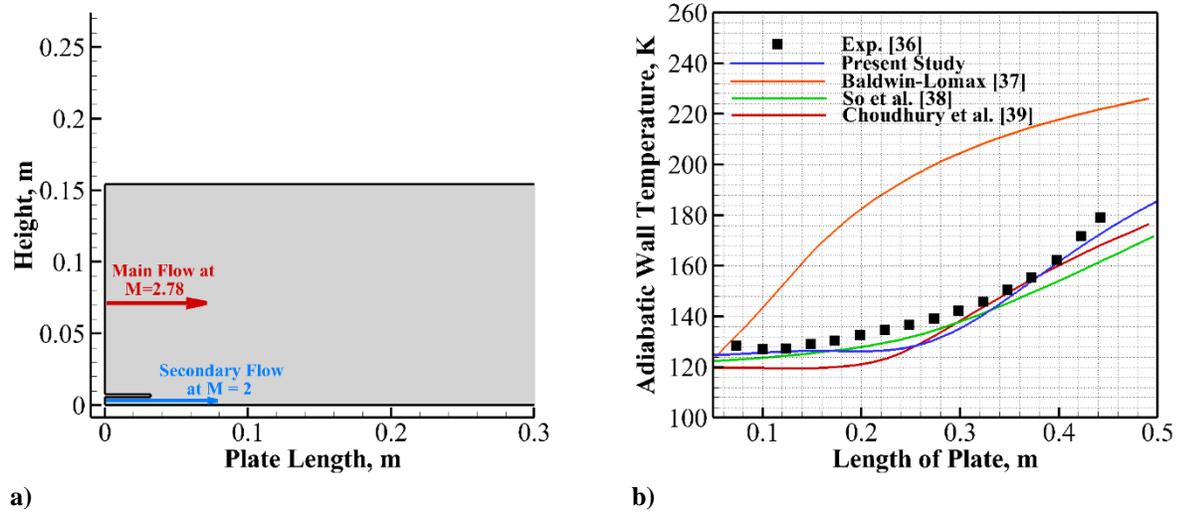

**Fig. 7.** Validation study of film cooling modeling **a)** Schematic of the supersonic flow injection over a flat plate **b)** Comparison of the adiabatic wall temperature over the length of the plate with different previous studies.

### III.B. Flow Development and Operating Mode Transition

As the rocket starts from the ground level, the outside pressure during its ascent continuously drops, increasing the NPR. Due to its geometrical nature, the dual bell inherits the two operational modes viz high-altitude mode and low-altitude mode.

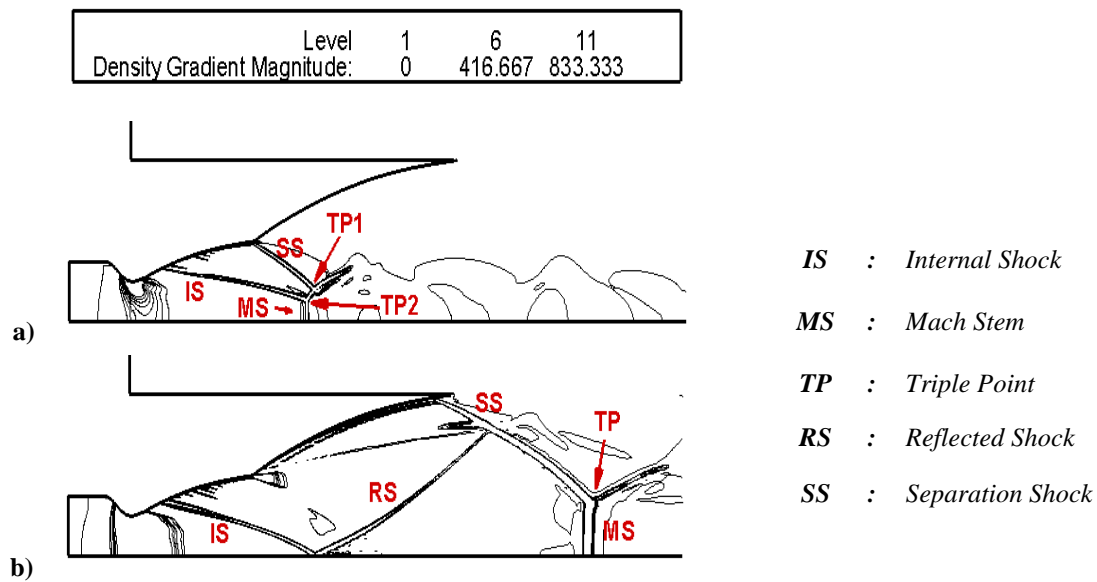

**Fig. 8.** Density gradient magnitude for **a)** low-altitude operation mode **b)** high-altitude operation mode.

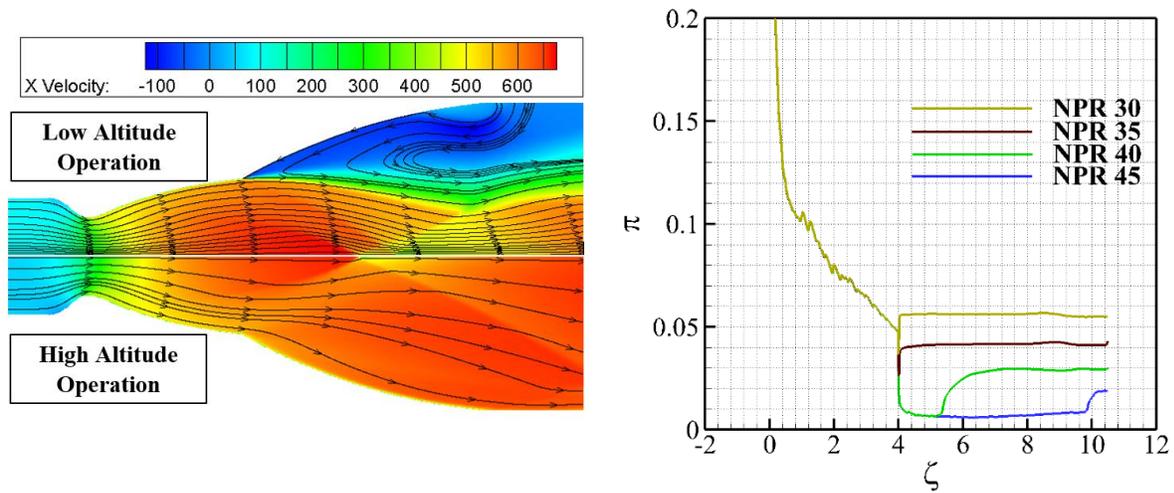

**a)** **b)**
**Fig. 9. a)** Axial velocity contour for different operation modes **b)** Wall pressure profiles for different NPRs corresponding to different operating modes

As observed for low-altitude operation mode, the flow separates from the inflection region at the end of the base nozzle. Fig. 8 confirms that this separation gives rise to an oblique separation shock that meets the Mach Stem (MS) formed upstream of the nozzle end, generating a triple point, TP1. The MS resembles the Type-2 Mach reflection [33-35] and occurs when the pressure increases radially from the axis, upstream of the MS. Figure 5 b) depicts the normalized wall pressure distribution ($p/p_c$), π, over the nozzle wall. While looking at Fig. 5 b), we notice that the pressure in the base nozzle decreases continuously until the start of the extension nozzle. From Fig. 5 b), one can also observe small oscillations in the wall pressure profile from $1 < \zeta < 3.5$. These small oscillations in the wall pressure profile are due to the formation of the weak shock waves on the nozzle wall due to the change in the wall contour of the base nozzle. These small weak shock waves can be seen in the density magnitude plot in Fig. 8. The flow separation at the inflection point increases the wall pressure. The flow from the ambient fills the extension nozzle resulting in circulation near the lip of the extension nozzle.

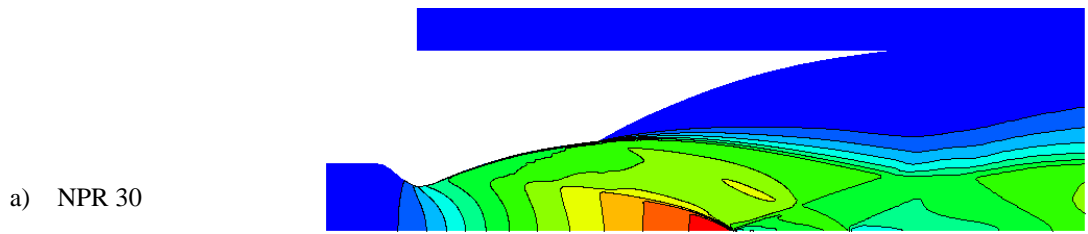

a) NPR 30

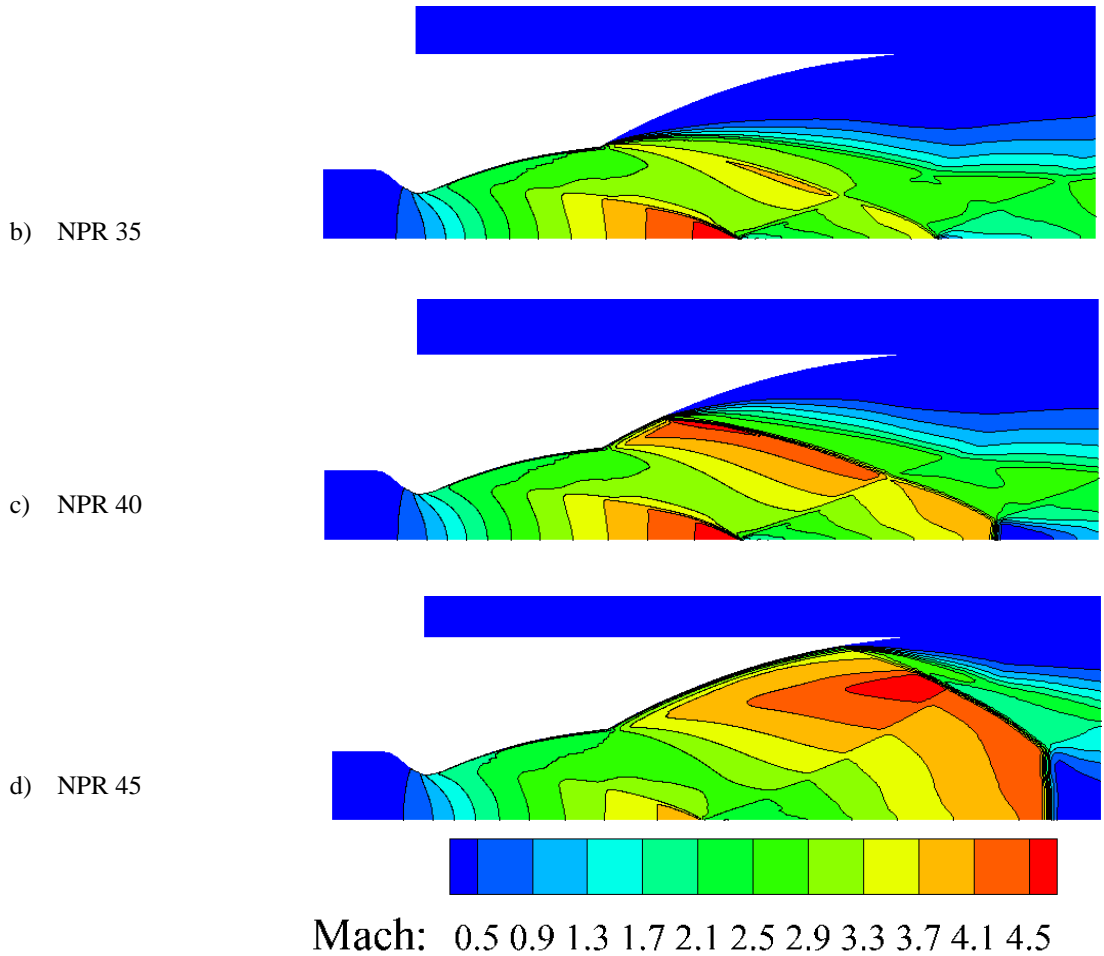

**Fig. 10.** Mach Number contours for different NPRs showing Operational Mode transition

On the other hand, for high-altitude operation mode (Fig. 9), the extension nozzle is full flowing with the main flow, and the flow separation occurs at the lip of the extension nozzle. As observed (Fig. 8 b), the separation shock meets the MS downstream of the nozzle end. This resembles the Type-1 Mach reflection [40-43] and occurs when the pressure decreases radially from the axis upstream of the MS. Unlike the low-altitude operation mode, the wall pressure further decreases after the inflection point (Fig. 5 b). Thus from Fig. 10, at the low altitude mode on NPR 30, as we keep on increasing the NPR from NPR 30 onwards, the flow separation shifts from the inflection point to the extension nozzle as can be observed in Fig. 10 c) at NPR 40, but remains close to the inflection point. On further increase in the NPR, the flow separation suddenly shifts near the end of the extension nozzle, as can be seen in Fig. 10 d) for NPR 45. Thus, the value of NPR, after which the flow separation points shifts near the end of the extension nozzle (as NPR 40 in our case) is known as 'transition NPR' for the dual bell nozzle, and the Mach number contours for different NPRs can be seen in Fig. 10. We can also observe from Fig. 9 b) that the separation point transits from the inflection point to the end of the extension nozzle with the increase in the NPR in the dual bell nozzle.

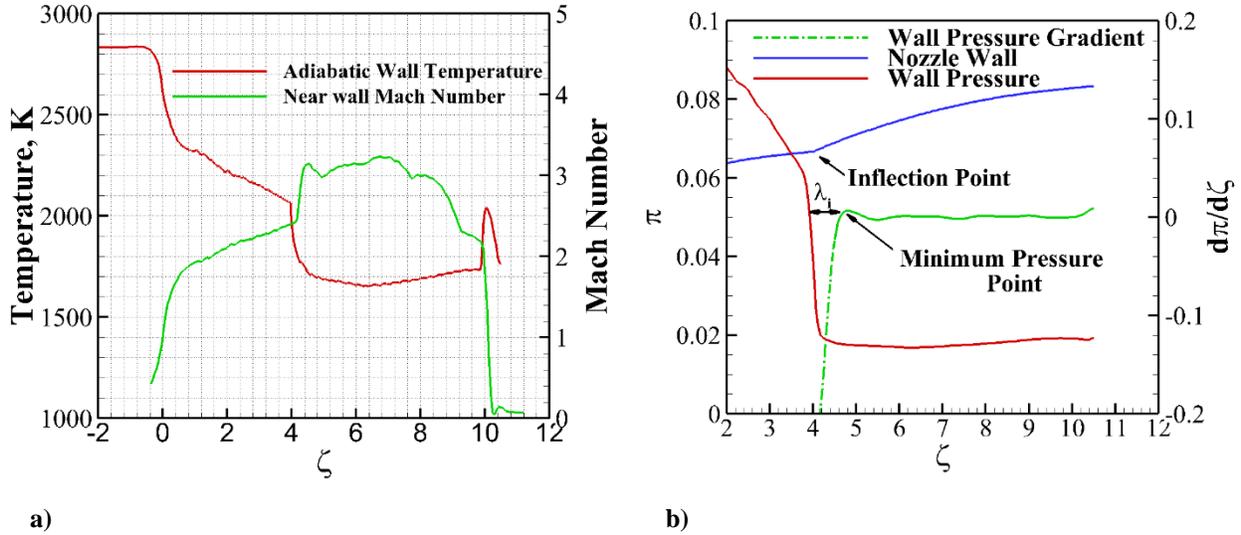

**Fig. 11. a)** Temperature profiles and **b)** Wall pressure behavior in the extension nozzle of the dual bell nozzle and definition of inflection region $\lambda_i$.

Figure 11 presents the adiabatic wall temperature distribution, Mach number distribution near the nozzle wall, and the wall pressure behavior for the dual bell nozzle at high-altitude operation mode without the film cooling. After the inflection point ($\zeta = 4$) in Fig. 11 a), we observe that the Mach number rapidly increases while the wall temperature shows a sudden decrease at the inflection point and then starts to increase again at a lower rate, which confirms the presence of the expansion fan at the inflection region. Furthermore, from Fig. 11 b), we can see that the inflection point is not the minimum pressure point on the nozzle wall. The region spanning from the nozzle inflection point to the minimum pressure point in high altitude mode is defined as 'inflection-region', denoted by $\lambda_i$ [7].

### III.C. Transition by a Sequence of Steady-State Computations

The literature of the dual bell nozzle suggests that the flow separation occurs at different NPR values during the case of up-ramping and down-ramping of the pressure feed, giving rise to the hysteresis phenomena [15-17]. Following the work of Martelli et al. [44], we have performed sequential computations on the steady-state by increasing the inlet pressure and keeping the fixed pressure at the outlet. The step size between the simulations is set to 2 bar. Figure 12 reports the variation in the location of the separation point obtained from the sequential steady-state simulations at different values of NPRs. During the low altitude operation (at low NPRs), the nozzle flow separates near the inflection point ($\zeta_{sep} \sim 4$). At approximately NPR 40 for up-ramping of pressure feed, the flow separation point starts shifting to extension nozzle from the inflection point. Further increase in the NPR from NPR 40 results in the shifting of the flow separation point near nozzle exit in the extension nozzle. This corresponds to the high-altitude operating mode of the dual bell nozzle. During the down-ramping of the pressure feed, the flow separates near the nozzle exit till NPR 38. At approximately NPR 35, the separation point shifts to the inflection point resulting in the transition from the high-altitude operation mode to low-altitude operation mode. Thus, the sequential steady-state simulations well reproduce the transition phenomena.

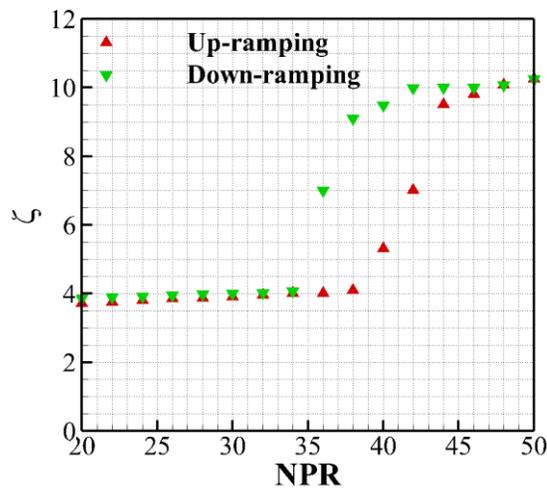

**Fig. 12.** Separation location obtained with the sequential steady-state simulations for different NPRs.

### III.D. Parametric Study of Film Cooling by Supersonic Injection

Previous studies of the dual bell nozzle [12-17] report that the dual bell nozzle offers a one-step transition from the sea level mode to the altitude mode, while the flow separates near the end of the nozzle extension at this transition NPR. However, for maximum thrust, it should separate at the end of the nozzle. The Pressure distribution for NPR 45 (Fig. 10) confirms that the flow separates near the end of the extension nozzle, and for the high value of NPR (say NPR 90, Fig. 5 b)), the flow separation is at the nozzle end. Thus, the full-flowing extension nozzle occurs at a higher NPR as opposed to the transition NPR. The flow should not separate before the end of the nozzle to obtain maximum thrust from the nozzle during the high-altitude operation mode. A secondary injection, which is used to cool the nozzle, has a significant impact on the movement of the separation point [12, 20-23, 39, 45]. In the next section of the paper, we have extensively documented the effect of the film cooling on the separation point during the flow transition.

In the previous studies on the film cooling of a dual bell nozzle, the authors have randomly selected the injection location for the secondary flow without giving any justification [12, 20-23]. Therefore, we have performed a systematic parametric study for the supersonic cooling inside the dual bell nozzle at the transition NPR, i.e. NPR 40, by injecting the secondary flow at different axial locations on the nozzle wall to obtain the most promising and effective injection location. The main flow inside the nozzle is developed with an inlet temperature of 2842 K and a pressure of 40 bar. Then, the secondary flow is injected with a temperature of 500 K, having a Mach number $M_f = 2$ with a film cooling slot height of 1 mm. The conclusion is made by studying the effect of the secondary injection on the pressure profile and the temperature distribution, as well as the mixing characteristics, assessed through the velocity profiles near the nozzle wall at different axial locations after the injection point.

*<u>Effect of Injection Location on Wall Temperature and Pressure</u>*

The secondary flow with Mach number $M_f = 2$ and temperature 500 K is injected parallel to the nozzle axis at three different axial locations on the nozzle wall viz. at the nozzle throat ($\zeta = 0$), at the base nozzle ($\zeta = 2$), and at the inflection point ($\zeta = 4$), respectively. Figure 13 illustrates the temperature contours corresponding to the three injection locations. From Fig. 13, we can observe that for inflection point injection, the secondary flow penetrates more into the hot main flow resulting in effective and relatively better cooling of the extension nozzle as compared to the other two. Further, Fig. 14 depicts the normalized wall pressure (Л) variation over the nozzle wall. As observed, the wall pressure for all three injections continuously decreases to the point where the flow separates from the nozzle wall. Notably, for the inflection point injection, the pressure starts increasing at the higher value of $\zeta$; while for the injection at the nozzle throat and the base nozzle, the wall pressure starts to increase at the approximately same value of $\zeta$, lower than the former one.

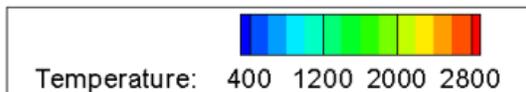

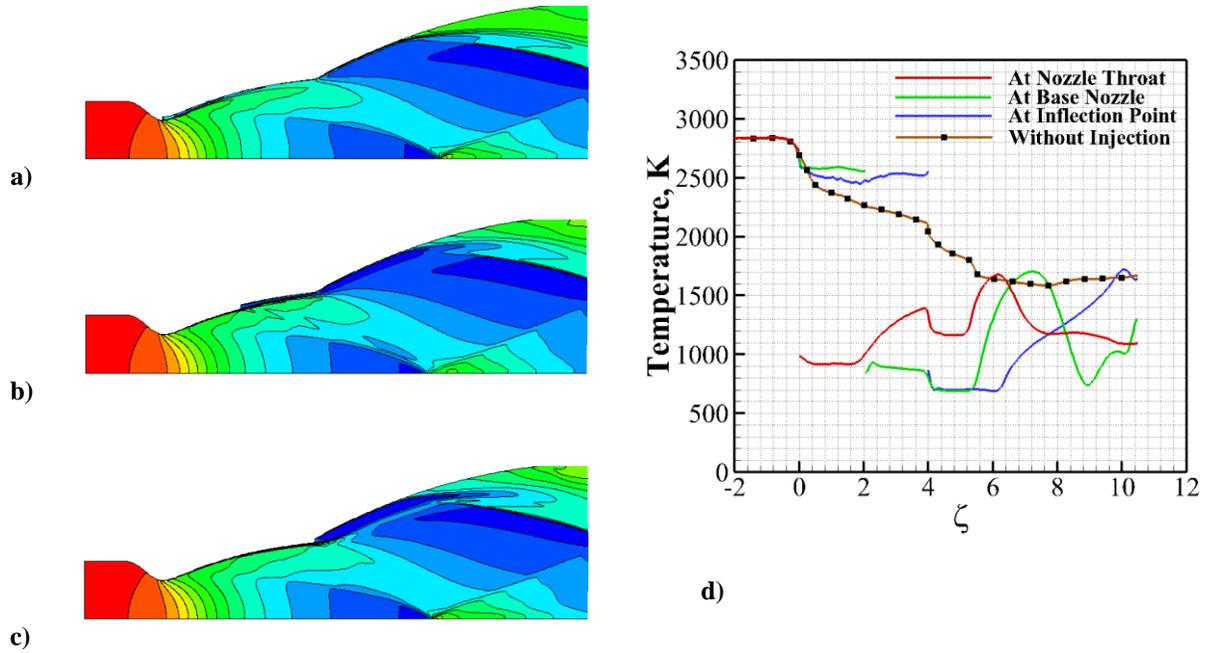

**Fig. 13.** Temperature contours for film cooling at NPR 40 for injection at **a)** nozzle throat ($\zeta = 0$) **b)** Base nozzle ($\zeta = 2$) **c)** Inflection region ($\zeta = 4$) **d)** Temperature distribution over nozzle wall.

Table 6 summarizes the location of the flow separation point on the nozzle wall for three injection locations. Thus, a higher value of $\zeta_{sep}$ (Non-dimensional separation point, $x_{sep}/r_{th}$ ) confirms that the flow remains more attached to the nozzle wall for the injection at the inflection point. This also results in more thrust as compared to the other two modes of operations.

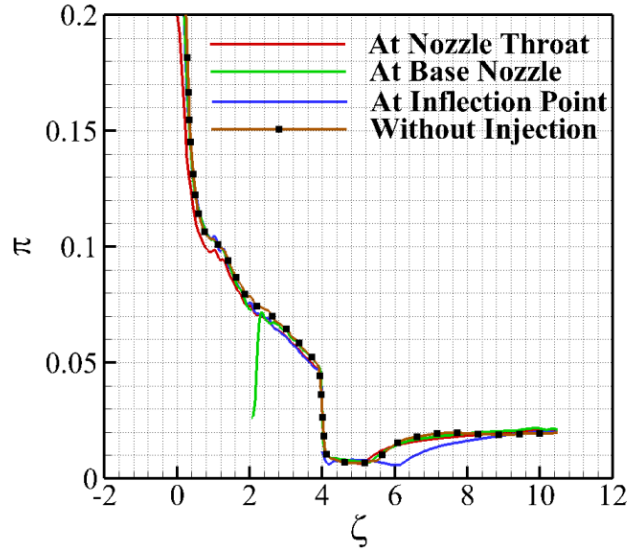

**Fig. 14.** Effect of the injection location for high-altitude operation (at NPR 40) on normalized Wall Pressure plot (Л)

**Table 6**
Separation point location for injection at different locations

| Sr. No. | Secondary Stream Injection location | Separation Point ($\zeta_{sep}$) |
|---|---|---|
| 1. | Nozzle Throat | 5.03 |
| 2. | Base Nozzle | 5.21 |
| 3. | Inflection Point | 6.12 |

*Effect of Injection location on Boundary Layer Profile and Thermal Profile*

To understand the mixing of the cold and hot flow inside the dual bell nozzle, we have assessed the boundary layer profiles as well as the near-wall temperature profiles at six different axial locations downstream of the secondary-flow-inlet, $\zeta_f$. Figure 15 shows the profiles mentioned above, corresponding to three different injection locations. The radial distance from the axis of the symmetry is normalized with the nozzle contour radius at that axial station, $y_w$. A quantity of 0.5 Mach translates each Mach number profile while each temperature profile is translated by 500 K for clarity reasons.

In the case of injection at the nozzle throat (Figures 15 a) and b)), the Mach number profiles and near-wall temperature profiles indicate the mixing of the secondary flow with the main flow. From Fig. 15 b), we observe that at $\zeta - \zeta_f = 2$, the secondary flow, and the main flow evolve to merge into a uniform temperature along the radial direction. In the case of injection at the base nozzle (Figures 15 c) and d)), the development of the Mach number profile and the near-wall temperature profile are different from the previous one. Here (Fig. 15 d), we can see that the secondary flow gradually mixes with the main flow, and uniformity in the temperature distribution starts to take place.

Moving ahead to Figures 15 e) and f), the Mach number profiles and the near-wall temperature profiles exhibit significant differences due to injection at the inflection point. As observed in Fig. 15 e), the spreading rate of the secondary flow is high, and the velocity profile rapidly evolves to merge into a uniform velocity along the radial direction (at approximately, $\zeta - \zeta_f = 1$). Also, no radial temperature gradient exists just after the injection inlet (Fig. 15 f). This is due to the relatively low temperature at the inflection region as compared to the other upstream injection locations. Further, the presence of an expansion fan at the inflection region further results in cooling in this region, which helps in making the temperature field uniform more quickly. This further supports the acceptance of the inflection point as the most effective location for the injection of the secondary stream.

### *Effect of Mach Number of Secondary flow injection on Wall Pressure, Wall Temperature, and Boundary Layer Profile*

Moreover, the injection of the secondary flow at the inflection point is carried out at four different Mach numbers to understand the effect of Mach number ($M_f$). The Mach number is varied by changing the injection velocity and keeping the inlet temperature constant. Further, the momentum flux ratio is also kept constant by varying the injection pressure for all four cases. Table 7 reports the corresponding pressure used for each case. The observations of the normalized wall pressure (Л) and wall temperature distribution over the nozzle wall provide the effect of Mach number. Figure 16 a) illustrates the boundary layer profile near the secondary flow inlet at NPR 40, while Figures 16 b) and c) depict the wall pressure distribution normalized with the combustion chamber pressure ($P_c$) and wall temperature distribution, respectively. We can observe that the Mach number of the injected secondary flow ($M_f$) has a significant effect on the nozzle wall pressure distribution. Figure 16 b) indicates that for Mach number $M_f = 1$, the flow separation moves upstream into the nozzle, resulting in the flow separation at the inflection point.

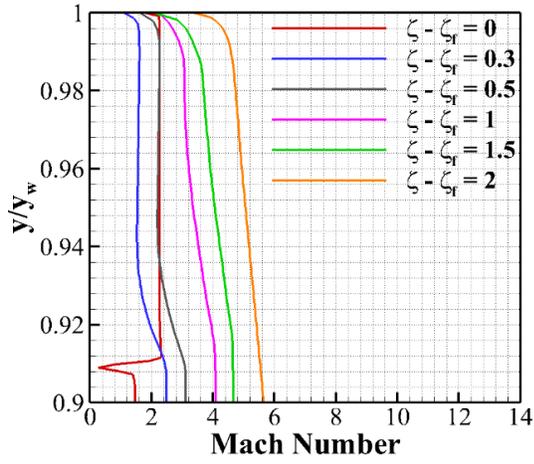
a)

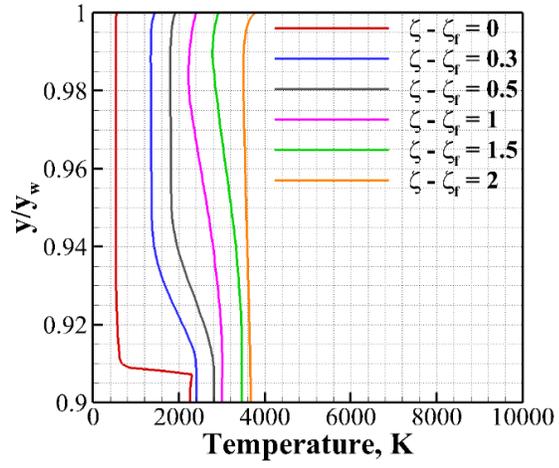
b)

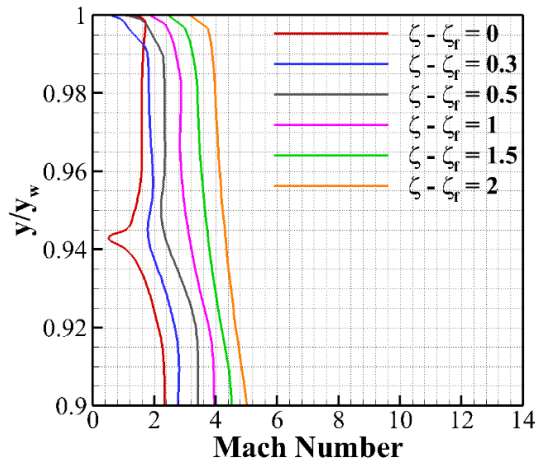
c)

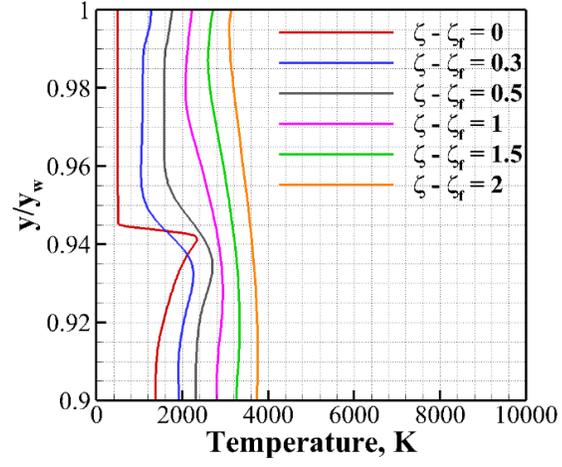
d)

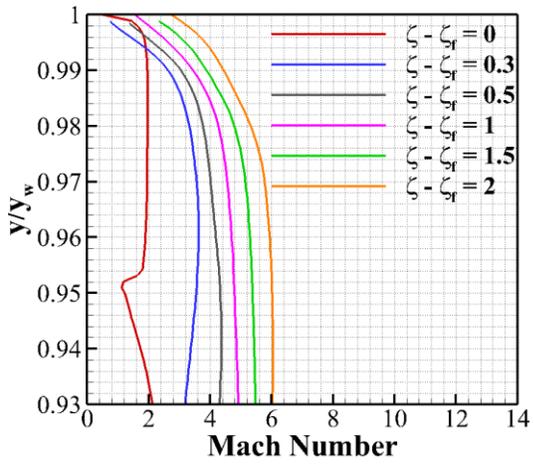
e)

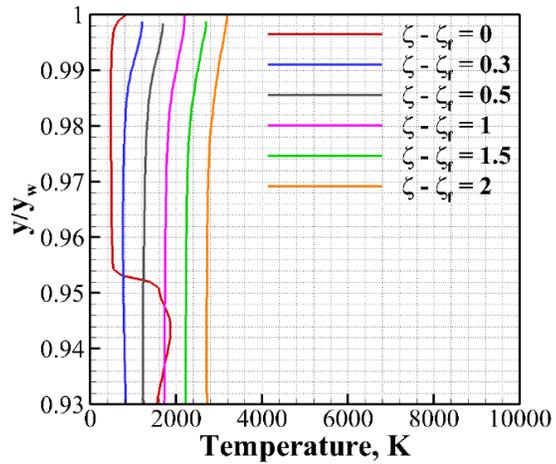
f)

**Fig. 15.** The boundary layer and temperature profiles at different axial locations for injection at **[a), b)]** Nozzle Throat ($\zeta = 0$)**, [c), d)]** Base nozzle ($\zeta = 2$) and **[e), f)]** Inflection Region ($\zeta = 4$)

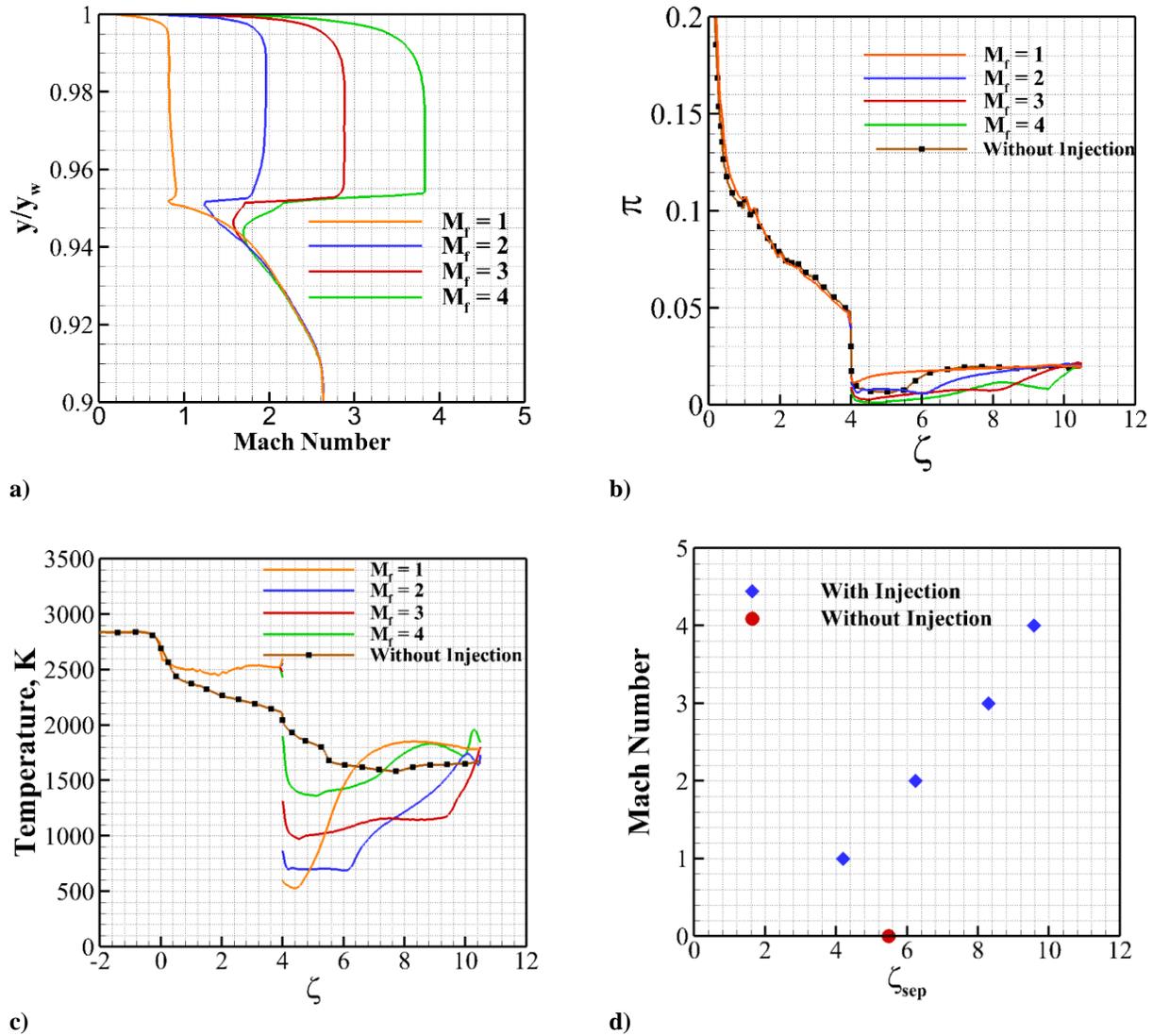

**Fig. 16.** Effect of the Mach number of Secondary Injection at the inflection point **a)** Mach number profile, **b)** Normalized wall pressure distribution (Π), **c)** Wall temperature distribution, **d)** Mach number vs. separation point

To maintain the constant mass flux ratio of the coolant and the main flow, the pressure at the inlet of the secondary injection is prescribed as 0.8 MPa (for $M_f = 1$ case), which is considerably higher than the pressure corresponding to the other Mach numbers of the coolant. This high value of coolant pressure imposes a relatively higher adverse pressure gradient, forcing the separation to occur at the inflection point itself for the case of Mach 1 coolant injection. At Mach number $M_f = 2$, the flow remains attached to the nozzle wall until $\zeta \sim 6$. The coolant jet with higher Mach no. (at the same temperature) contains higher momentum, which allows it to overcome adverse pressure gradients resulting in late separation on the nozzle wall. With further increase in the Mach number of the secondary flow, although the separation point further shifts downstream into the extension nozzle near the nozzle exit ($\zeta \sim 10$), the pressure oscillations start to increase over the nozzle wall, which may further result in the high values of the undesirable side loads. Figure 16 d) shows the movement of separation point with an increase in Mach number of the secondary flow ($M_f$). Thus, we can see that we can also control the transition phenomena with the increasing or decreasing the Mach number of the injected secondary flow.

**Table 7**
Properties specified at the inlet of the secondary flow

| Case. No. | Coolant Mach Number | Coolant Inlet Pressure ($\times 10^5$ Pa) | Coolant Inlet Temperature (K) |
|---|---|---|---|
| 1. | 1 | 8 | 500 |
| 2. | 2 | 2 | 500 |
| 3. | 3 | 0.88 | 500 |
| 4. | 4 | 0.50 | 500 |

## IV. Conclusion

The analysis carried out in this study assesses the flow features related to the dual bell nozzle. The numerical results are initially verified and then validated with the experimental results obtained by Tomita et al. [6]. The structure of the Mach Stem is compared with the findings of Nasuti et al. [40] by studying the density gradient magnitude plots. The Mach Stem for low-altitude operation resembles a Type-2 Mach reflection while the Mach Stem for high-altitude operation resembles a Type-1 Mach reflection. The transition from low-altitude operation mode to high-altitude operation mode is also numerically re-produced by the sequence of the steady-state simulations.

Further, the secondary flow is injected in the dual bell nozzle to assess the film cooling process. The injection location has a significant effect on the movement of the separation point on the nozzle wall. Moreover, it is found that the inflection point is the most effective location for the injection of the secondary flow while examining the wall pressure profiles, temperature profiles, and the boundary layer profiles. Finally, we report the effect of the Mach number of the secondary flow. As observed, the separation point shifts downstream with the increase in the Mach number of the secondary flow (except the Mach 1), which results in a higher amount of thrust. Thus, with the help of the Mach number of the secondary flow, we can actively control the shifting of the separation point on the nozzle wall or, in turn, we can control the transition phenomena in the dual bell nozzles. Although, this would result in the increased pressure fluctuations over the nozzle wall and thereby contributing to the undesired side-loads. The appearance of side-loads also poses a limit on increasing the Mach number of the secondary flow beyond a specific value. One may carry out a transient study to optimize other different parameters like temperature, Mach number, and the mass ratio of the secondary flow, which remains as a scope for further studies.

## Conflict of interest statement

We wish to confirm that there are no known conflicts of interest associated with this publication.

## Acknowledgments

Financial support for this research is provided through IITK-Space Technology Cell (STC). The authors would also like to acknowledge the High-Performance Computing (HPC) Facility at IIT Kanpur (www.iitk.ac.in/cc).